# Incident-angle dependence of deformation characteristics of aluminum surface under low-energy xenon-ion impact

Cheng Zhang,[a,1], Jiang Zhou [a], Weihua Xie [a], Jiacong Yin [a], Hongjiao Zhao [b]

[a] *Institute of Telecommunication and Navigation Satellites, CAST, Beijing, 100094, China*

[b] *School of Aerospace Engineering, Key Laboratory for Thermal Science and Power Engineering of Ministry of Education, Tsinghua University, Beijing 100084, China*

**Abstract**

Ion thruster is a revolution technology with potential applications in space mission but the thruster's operation lifetime is limited by the sputtering from thruster components. In this work, molecular dynamic simulations are performed to explore the dependence of deformation characteristics of an aluminum surface on incident angle and kinetic energy under low-energy xenon-ion impact. The fraction of non-12-coordinated atoms is used to quantitatively characterize the microstructural evolution and defect density levels. It is found that defect density level has a linear relation with incident energy, and there exists a critical incident angle around 20°, at which the aluminum surface has the maximum defect density level. In addition, a collision model is developed to theoretically reveal the physical mechanisms behind the dependence. Our findings may helpful in developing long endurance electric propulsion devices for practical applications.

---

[1] Author to whom correspondence should be addressed. Electronic mail: zhangcheng4@126.com

# 1. Introduction

Electric propulsion, a key and revolutionary technology for the new generations of satellites, has attracted considerable attention in recent years due to its outstanding advantages such as superior fuel efficiency, unprecedented specific impulse, high thrust precision, low costs, and long endurance [1–4]. These properties make it promising technology for potential applications in a broad range of fields such as orbit correction and deep space exploration [5,6].

To generate thrust, electric propulsion electrostatically accelerate xenon ions through molybdenum grids, which becomes a common form of spacecraft propulsion. However, when the ions fly outside the thruster, some collisions may happen with the screens because not all ions move out the thruster cleanly [7]. Due to the kinetic energy of energetic ions, these collisions will induce sputtering and further erosion. Then damage caused by impacts to the solid surfaces of thruster may results in the formation of defects such as nanovoid, adatoms and so on, which limits the thruster's operation lifetime [8,9].

A variety of theoretical methods such as empirical model [10,11] and analytical method [12,13] have been developed toward exploring the sputtering. However, these models require parameters from experiments and are unable to provide satisfactory description for low-energy ions [7,14]. Molecular dynamics (MD) simulations can be a powerful tool to explore electric propulsion with better accuracy for low-energy ion-crystal interactions. For instance, Elliott *et al.* [7] measured sputter yields of twelve different elemental targets under xenon-ion bombardment at normal incidence angles by performing atomistic simulations. Their results have shown the details of the low-energy sputter yield as a function of ion incidence angle, the kinetic energy distributions and probability density functions for the trajectories of sputtered atoms. Zhou *et al*. [8] utilized atomistic simulations to study the low-energy (50-300 eV) normal incident xenon-ion sputtering on low index nickel surfaces. Their results indicate the existence of a sputtering threshold ion energy of ~25 eV, below which no sputtering occurs. Gen *et al.* [14] investigated the ion-wall reflection characteristics by combining MD simulations and a practical model, and found that the surface roughness effects should not be ignored in the extended Cercignani-Lampis-Lord model. Henriksson *et al.* [15] performed sputtering simulations to demonstrate the size distribution and temperature of clusters produced by 15 keV xenon-ion impacts on silver and 20 keV xenon-ion impacts on gold.

Previous research on interaction between energetic particles and solid surfaces mainly focus on sputter yields. To date, the deformation and damage characteristics of surface induced by low-energy ion-crystal interactions is still open to further investigations. In the present work, a MD approach is used to investigate the detailed deformation behavior of an aluminum surface under the bombardment of xenon ion (Xe). The incident angle and kinetic energy of Xe are considered to characterize the deformation. Moreover, insights from these simulations are used to develop an analytical theory to reveal the underlying physical mechanisms behind incident-angle and kinetic-energy dependence.

## 2. Methods

Our simulation method is illustrated in Fig. 1. A xenon ion is placed above an aluminum substrate, which has a structure of face-centered cubic (FCC) single crystal with a lattice constant of 4.05 Å and a size of $20\times20\times10$ crystal cells. The Cartesian *x*-, *y*-, and *z*-axes are taken along the [100], [010], and [001] crystallographic directions, respectively. Periodic boundary conditions are imposed in *x*- and *y*-axis directions, and free boundary conditions are applied to the boundaries of *z*-axis. Aluminum atoms in the bottom two crystal cells height of substrate are fixed. The initial distance between the xenon ion and the center of the substrate surface is five crystal lattices.

Before impact, the system is allowed to relaxed to reach its minimum energy state by using conjugate gradient method. Then a Nose-Hoover thermostat is employed to thermally equilibrate the system for 2 ps. In order to avoid noise from thermal effects, simulations are run at temperature of 1 K. After relaxation, impact is applied by emitting Xe to the center of the aluminum substrate surface with a kinetic energy $E_k$ at a chosen incident angle *θ*. To investigate the dependence of deformation behaviors of the aluminum substrate on kinetic energy and incident angle, the initial kinetic energy of Xe ranges from 10 eV to 100 eV and the incident angle changes from 0° to 90°. During simulations, the common neighbor analysis (CNA) [16] is adopted to identify atoms in particular environment such as FCC, hexagonal close-packed (HCP), body-centered cubic (BCC), icosahedral and non-12-coordinated environment in the aluminum substrate.

We perform MD simulations using LAMMPS package [17]. The embedded-atoms method (EAM) potential developed by Zhou *et al.* [18] is applied to describe the interaction between Al atoms. In the EAM model, the total energy $E^{\text{EAM}}$ is given by the following equation

$$E^{\text{EAM}} = \frac{1}{2} \sum_{j \neq i} \phi_{ij}(r_{ij}) + \sum_i F_i(\rho_i), \qquad (1)$$

where $r_{ij}$ is the distance between atoms *i* and *j*, $\phi_{ij}$ is the pair energy, and $F_i$ is the embedding energy associated with embedding an atom *i* into a local site with an electron density $\rho_i$. This potential is well fitted to basic material properties such as lattice constants, elastic constants, bulk moduli, sublimation energies and vacancy formation energies [19].

Ziegler-Biersack-Littmark (ZBL) universal potentials are used for the Xe–Al interactions. In the ZBL model, the potential energy between atoms $i$ and $j$ at a distance $r_{ij}$ can be expressed as

$$E_{ij}^{ZBL} = \frac{1}{4\pi\epsilon_0} \frac{Z_i Z_j e_0^2}{r_{ij}} \phi^{ZBL}(r_{ij}/a) + S(r_{ij})$$

$$a = \frac{0.4685}{Z_i^{0.23} + Z_j^{0.23}} \quad , \quad (2)$$

$$\phi^{ZBL}(x) = \sum_{i=1}^{4} A_i e^{-B_i x}$$

where $\epsilon_0$ is the electrical permittivity of vacuum, $e_0$ is the electron charge, $Z_i$ and $Z_j$ are the nuclear charges of the two atoms, $a$ is the screening length, and $\phi^{ZBL}$ is the screening function with coefficients **A**=[0.18175, 0.50986, 0.28022, 0.02817], **B**=[3.19980, −0.94229, −0.40290, −0.20162]. $S(r_{ij})$ is an additional switching function which ramps the potential and force smoothly to zero between the inner and outer cutoff distances.

$$S(r) = \begin{cases} 0, & r \geq r_2 \\ \frac{S_a}{3}(r-r_1)^3 + \frac{S_b}{4}(r-r_1)^4, & r_1 < r < r_2 \\ S_c, & r \leq r_1 \end{cases} \quad (3)$$

where

$$\begin{aligned} S_a &= \left[-3E'(r_2) + (r_2-r_1)E''(r_2)\right]/(r_2-r_1)^2 \\ S_b &= \left[2E'(r_2) + (r_2-r_1)E''(r_2)\right]/(r_2-r_1)^3 \\ S_c &= -E(r_2) + (r_2-r_1)E'(r_2)/2 - (r_2-r_1)^2 E''(r_2)/12 \end{aligned} \quad , \quad (4)$$

$r_1$ and $r_2$ are the inner cutoff and outer cutoff distance, respectively. The boundary conditions applied to the smoothing function are as follows: $S'(r_1) = S''(r_1) = 0$, $S(r_2) = -E(r_2)$, $S'(r_2) = -E'(r_2)$ and $S''(r_2) = -E''(r_2)$, where $E(r)$ is the first term of the $E^{ZBL}$. Single and double primes denote first and second derivatives with respect to $r$, respectively. The coefficients $S_a$, $S_b$, and $S_c$ are computed to perform the shifting and smoothing. The inner cutoff was set at 3 Å, and the outer cutoff was set at 4 Å.

## 3. Results and Discussion

### 3.1 Deformation behaviors and physical mechanisms

To investigate whether the deformation features of aluminum substrate depends on the incident angle, we first study two cases with the same kinetic energy of 80 eV but different incident angles, narrow-angle 10° and wide-angle 70°. Figures 2 and 3 capture sequences of snapshots of the microstructural evolution of these two cases. Side elevations from *x*- and *y*-axis are illustrated in the first and second rows in each figure, respectively. For a clearer recognizing, perfect FCC atoms are not shown for viewing. It is observed that during impact there are only non-12-coordinated atoms, and no HCP, BCC or icosahedral atoms appear. The non-12-coordinated atoms usually appear in the free surface region or in the dislocation core [20]. As a result, the microstructural evolution and defect density levels can be quantitatively characterized by the fraction of new non-12-coordinated atoms inside the aluminum substrate (FNA), which is defined as $\eta = (N_n - N_s)/N_a$, where $N_n$ is the number of non-12-coordinated atoms, $N_s$ is the number of atoms on the surface, and $N_a$ is the total number of atoms in the substrate. The variations of FNA $\eta$ with time for cases with different incident angles are displayed in Fig. 4. For each case, FNA starts to increase from zero as Xe reaches the substrate surface, which moment is defined as the beginning of the curve (time zero).

A typical FNA-time curve for the case with a narrow incident angle 10° is shown as the black line in Fig.4, which can be characterized by the following three regimes of distinct deformation behaviors. In the first regime ($A_1C_1$), the FNA $\eta$ increases with time until the maximum FNA $\eta_m$=3.96% at Point $C_1$ is reached. As the time further increases, the FNA-time curve enters the second regime ($C_1E_1$), which exhibits a negative slope. In the third regime ($E_1F_1$), in which the time exceeds $t$=0.78 ps in the simulation, the FNA $\eta$ remains nearly the same.

Figure 2 presents a sequence of snapshots to display the microstructural evolution during deformation. When Xe reaches the surface of aluminum substrate, new non-12-coordinated atoms initially nucleate from surface (Fig. 2a). As the kinetic energy of Xe transferers to the substrate, the number of non-12-coordinated atoms pronounced increase (Fig. 2b) to the maximum FNA (Fig. 2c). It is noted that in Fig.2c, the length, width and depth of the region of non-12-coordinated atoms in $x$-, $y$- and $z$-axis are 34.5 Å, 36.6 Å and 30.6 Å, respectively. Beyond the peak value, FNA decreases as the time is further increased. As the velocity of Xe changes from downward to upward, some non-12-coordinated atoms transform back to FCC atoms (Fig. 2d) because the rapid release of stress caused by impact, resulting in the deceasing of FNA in curve (Point $D_1$). After the time exceeds a critical value of $t$=0.78 ps (Point $E_1$), no evident change is observed in the FNA (Fig. 2e), indicating that the distance between reflected Xe and the surface is so far away.

Compared to case with narrow incident angle, the bombardment in a wide incident angle shows less influence on the substrate. Considering the case with the kinetic energy of 80 eV and incident angle of 70°, the microstructural evolution during impact is illustrated in Figs. 3a-f, and the FNA $\eta$ as a function of time is plotted as the pink curve in Fig. 4. It appears that the FNA-time curve and the microstructural evolution are similar to the case with incident angle of 10°. However, the maximum value of FNA is $\eta_m$=1.04% (Point $C_2$), which is 73.4% less than that of narrow-angle 10° case. The size of non-12-coordinated atoms region in three dimensions at Point $C_2$ are 12.4 Å, 22.7 Å and 16.3 Å, only 35.9%, 62.0% and 53.3% of the narrow-angle 10° case, respectively.

*3.2 Incident-angle and kinetic energy dependence*

To discuss the influence of incident angle $\theta$ and kinetic energy $E_k$ of Xe to the substrate, the maximum FNA $\eta_m$ for each case with various $\theta$ and $E_k$ are calculated. Figure 5a shows $\eta_m$ as a function of incident angle of cases with kinetic energy of $E_k$= 80 eV, 50 eV and 20 eV. Taking the case with $E_k$= 80 eV as the example, it is seen that as the incident angle rotates from the direction perpendicular to substrate ($\theta$=0°) to the direction parallel to substrate ($\theta$=90°), the maximum FNA $\eta_m$ oscillatingly increases to 4.25% at an incident angle of 20°, then it continuously decreases to 0.19% at the incident angle of 80°. The aluminum substrate has the largest ratio $\eta_m = 4.25\%$ at $\theta$=20°, indicating that aluminum substrate suffers largest deformation induced from the impact at this incident angle.

Figure 5b illustrates the relationship between the maximum FNA $\eta_m$ and kinetic energy $E_k$ for cases with three different incident angles. It is demonstrated that $\eta_m$ and $E_k$ obey approximately linear relation. By linearly fitting the $\eta_m \sim E_k$ curves, the coefficients for cases with incident angle $\theta$=0°, 40° and 70° are 4.05, 3.57 and 1.43 eV$^{-1}$/10$^4$, respectively. The difference of coefficients between cases with $\theta$=40° and 70° is 4.46 times larger than that between cases with $\theta$=0° and 40°, means the deformation of substrate induced by impact is more sensitive to kinetic energy at the narrower incident angles.

Figure 6 shows the distribution of the maximum FNA $\eta_m$ for each case with various incident angles $\theta$ and kinetic energies $E_k$. It is observed that these cases can be divided into three groups according to $\eta_m$. In Group I, $\eta_m$ is less than 1.0% and the impact has slight influence to the substrate. In this Group, the incident angle is larger than 70° or the incident energy is less than 10 eV. In Group II, $\eta_m$ is between 1.0% and 4.0%. Cases in Group III have large $\eta_m$ which is more than 4.0%. Interestingly, the maximum $\eta_m$ is located at the zone $E_k \in (90,100)$ eV and $\theta \in (15°, 30°)$, at which the aluminum surface has the maximum defect density level, as indicated by the black dash box.

*3.3 Modeling the ion impact on surface*

In order to explain the dependence of maximum FNA $\eta_m$ on the incident angle $\theta$ and kinetic energy $E_k$, a theoretical model is proposed. As depicted in Fig. 7, Xe is considered as a non-deformable rigid particle, which strikes the substrate with a velocity $v$ at an incident angle $\theta$. After reflection, the velocity becomes $v'$ at an angle $\beta$. Collision process can be described by impulse equations as follow

$$\begin{cases} mv\sin\theta - \int_0^t F_f(t)dt = mv'\sin\beta \\ mv\cos\theta - \int_0^t F_N(t)dt = -mv'\cos\beta \end{cases}, \quad (5)$$

where $F_f(t)$ and $F_n(t)$ are the forces between particle and substrate in tangent and normal directions, respectively. For the sake of simplicity, we use $I_n = \int_0^t F_N(t)dt$ to represent the impulse in the normal direction and assume $F_f(t)=\mu F_n(t)$ during impact, where $\mu$ is the friction coefficient. Before and after the collision, the difference of kinetic energy can be expressed as,

$$E_\Delta = v(\mu\sin\theta + \cos\theta)I_n - \frac{1+\mu^2}{2m}I_n^2. \quad (6)$$

To measure of how much kinetic energy remains after the collision of two bodies, previous collision models usually use the coefficient of restitution $e$ [21–23], which is defined as the ratio of the velocity components along the normal plane of contact after and before the collision. Here, the coefficient of restitution is $e = \dfrac{v'\cos\beta}{v\cos\theta}$, so the impulse in the normal direction can be expressed as

$$I_n = (1+e)mv\cos\theta. \tag{7}$$

Using the Eq. (7), $E_\Delta$ can be written as

$$E_\Delta = \left[2(1+e)(\mu\sin\theta+\cos\theta)\cos\theta-(1+e)^2(1+\mu^2)\cos^2\theta\right]E_k. \tag{8}$$

During impact, the kinetic energy of particle is transformed to the deformation energy of substrate, leading to the appearance of non-12-coordinated atoms. A larger kinetic energy difference $E_\Delta$ indicates a more maximum FNA $\eta_m$. According to Eq. (8), $E_\Delta$ has a linear relation with incident energy $E_k$. Thus, the maximum FNA $\eta_m$ shows a linear relation with incident energy $E_k$, agrees with the observation in Fig. 5b. To get the dependence of $E_\Delta$ on incident angle, the solutions for $\partial E_\Delta/\partial\theta=0$ is derived

$$\tan 2\theta_c = \frac{\mu}{1-(1+\mu^2)(1+e)/2}. \tag{9}$$

When the incident angle $\theta$ is less than the critical incident angle $\theta_c$, the energy $E_\Delta$ increases as the incident angle $\theta$ increases. After the incident angle exceeds $\theta_c$, the monotonicity of $E_\Delta$ inverses. According to Eq.(9), the critical incident angle $\theta_c$ is determined by two parameters, i.e., the friction coefficient $\mu$ and the restitution coefficient $e$. As can be seen from Fig. 7, varying the friction coefficient $\mu$ and the restitution coefficient $e$ only changes the value of critical incident angle $\theta_c$, but does not alter the shape of $E_\Delta/E_k$ curves, which has a fairly good agreement with the tendency in Fig. 5a. It is should be noted that in order to deduce the damage caused by impacts to the solid surfaces, one can choose the surface a material which has appropriate friction coefficient and restitution coefficient.

## 4. Conclusions

In summary, we have used MD simulations and erosion model analysis to investigate the deformation behaviors and underlying physical mechanisms of an aluminum surface under Xe impact with various incident angles and kinetic energies. The microstructural evolution and defect density levels are quantitatively characterized by the fraction of non-12-coordinated atoms inside the aluminum substrate. It is found that the defect density level obeys an approximate linear relation with respect to kinetic energy. The results also indicate that as the incident angle rotates from the direction perpendicular to substrate to the direction parallel to substrate, the defect density levels oscillatingly increases to the peak at a critical incident angle of 20°, then they continuously decrease. It is expected that the research could promote applications of ion thruster in engineering.

**Figures**

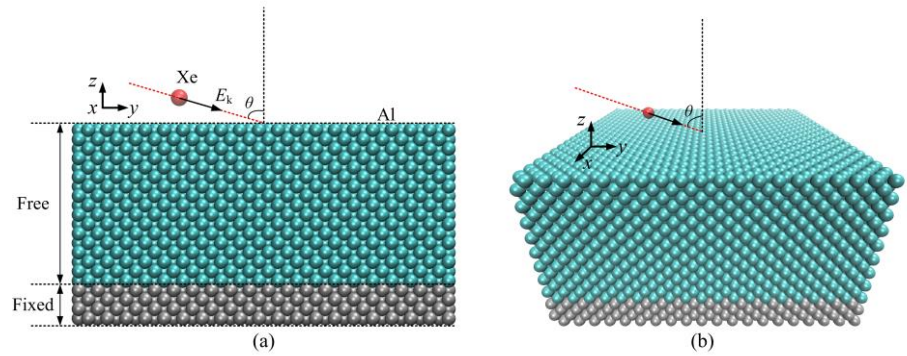

**Fig. 1 -** Simulation model. (a) lateral view and (b) oblique view of the atomic configuration. A xenon ion is placed above an aluminum substrate, which possesses {100} orientations in all three directions. Impact is applied by emitting Xe to the center of the aluminum substrate surface with a kinetic energy $E_k$ at an incident angle $\theta$.

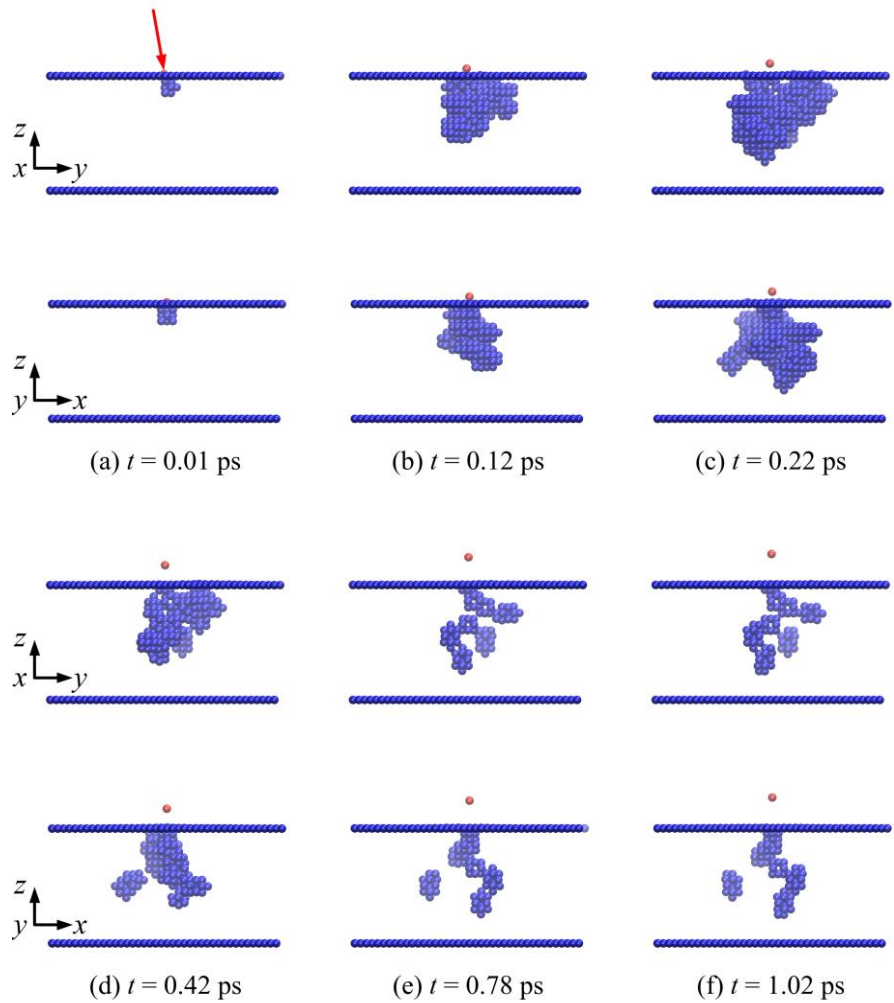

**Fig.2** – Microstructure Evolution of the case with kinetic energy of 80 eV and incident angle of 10°. Atomic configurations at time (a) 0.01 ps, (b) 0.12 ps, (c) 0.22ps, (d) 0.42 ps, (e) 0.78 ps and (f) 1.02ps, corresponding to the Points $A_1$, $B_1$, $C_1$, $D_1$, $E_1$ and $F_1$ in Fig. 4, respectively. Snapshots in top and bottom rows represent the lateral views from $x$- and $y$-axis. Atoms are colored according to the CNA method.

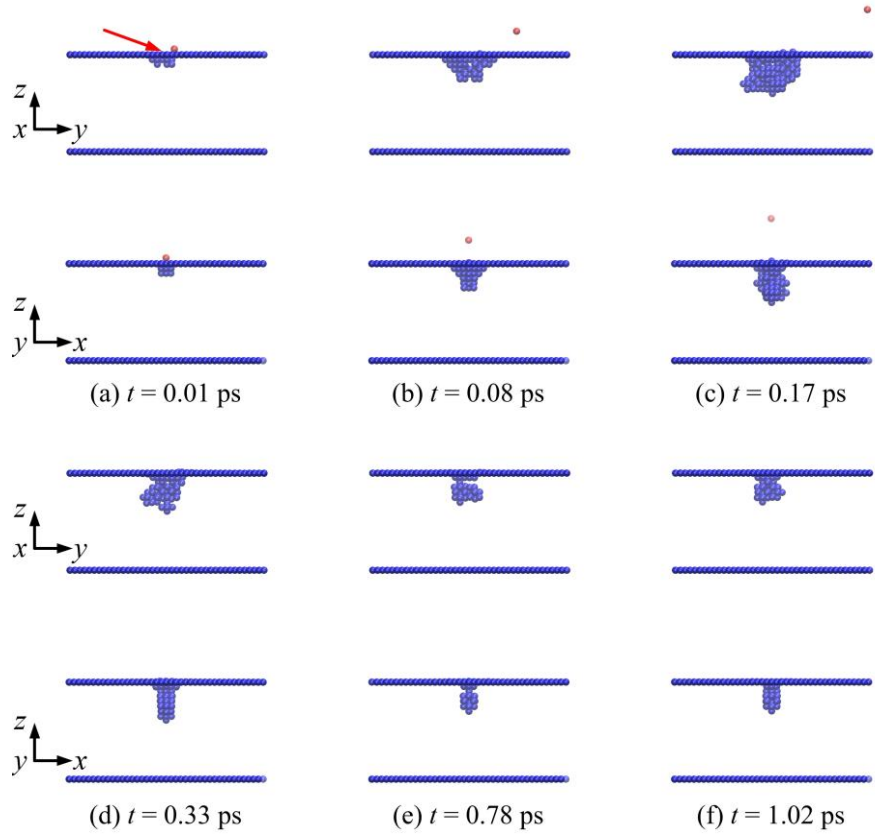

**Fig.3** – A sequence of snapshots for deformation of the case with kinetic energy of 80 eV and incident angle of 70°. Evolution of microstructure with increasing time at (a) 0.01 ps, (b) 0.08 ps, (c) 0.17ps, (d) 0.33 ps, (e) 0.78 ps and (f) 1.02ps, corresponding to the Points $A_2$, $B_2$, $C_2$, $D_2$, $E_2$ and $F_2$ in Fig. 4, respectively. Subfigures represented in top and bottom rows illustrate the lateral views from *x*- and *y*-axis. Atoms are colored according to the CNA method.

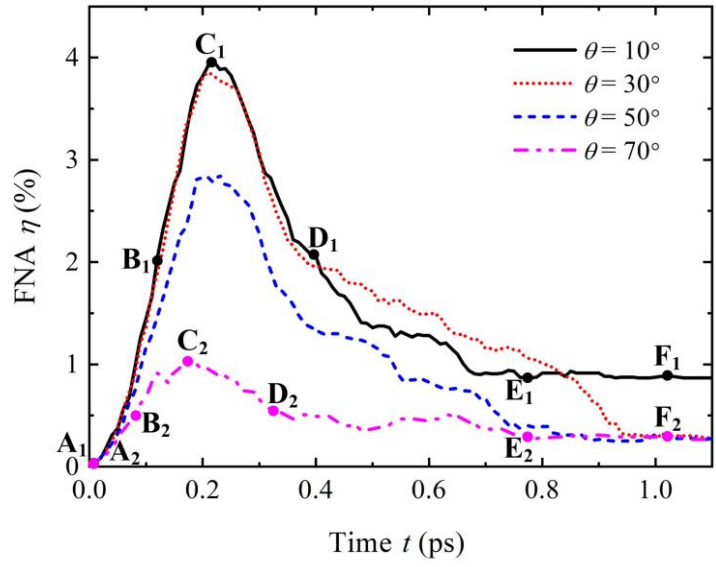

**Fig. 4** - FNA $\eta$ as a function of time for cases with different incident angles.

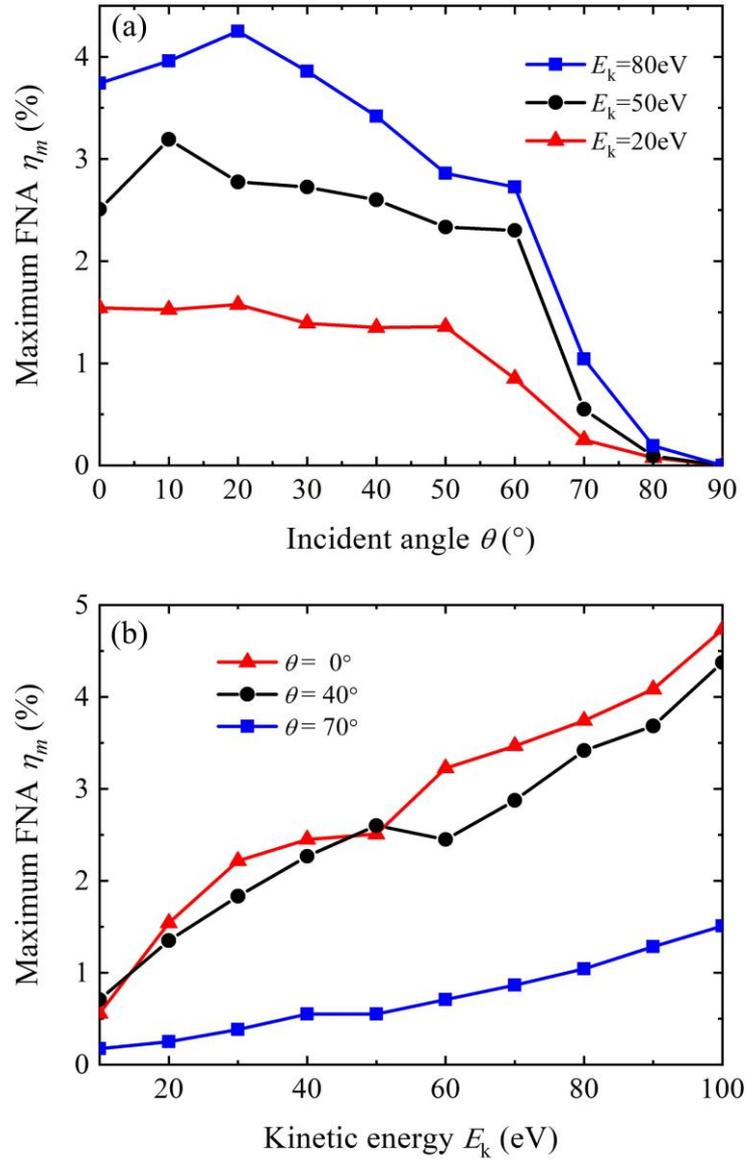

**Fig. 5** - Maximum FNA $\eta_m$ as functions of (a) incident angle $\theta$ and (b) kinetic energy $E_k$. As the incident angle rotates from the direction perpendicular to substrate ($\theta=0°$) to the direction parallel to substrate ($\theta=90°$), $\eta_m$ increases to the peak value at an critical incident angle, then it decreases. $\eta_m$ and $E_k$ obey an approximately linear relation.

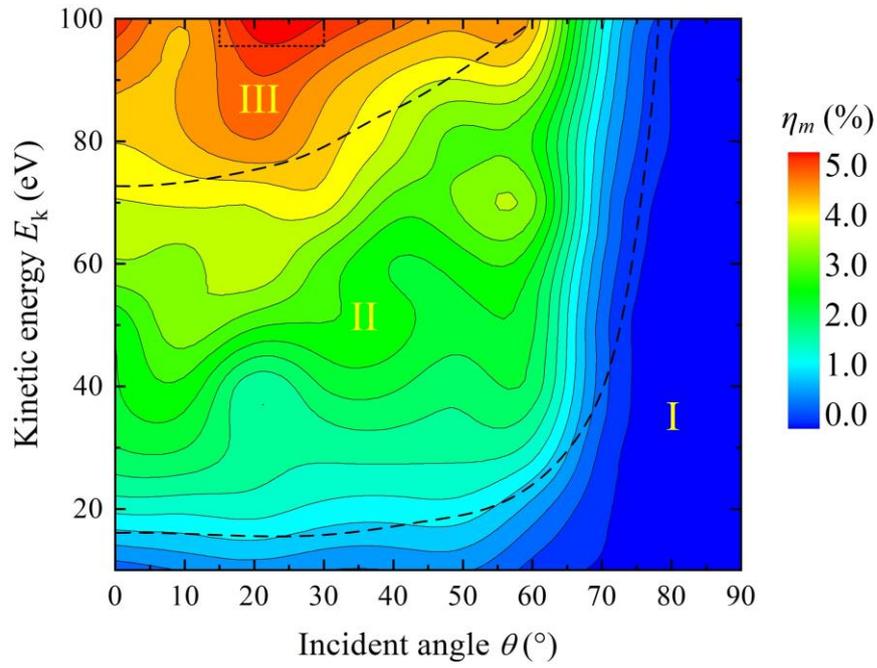

**Fig. 6** - Distribution of $\eta_m$ for each case with various incident angles $\theta$ and kinetic energies $E_k$. The distribution is divided into three groups according to the value of $\eta_m$: Group I, $\eta_m \leq 1\%$; Group II, $1\% < \eta_m < 4\%$; Group III, $\eta_m \geq 4\%$.

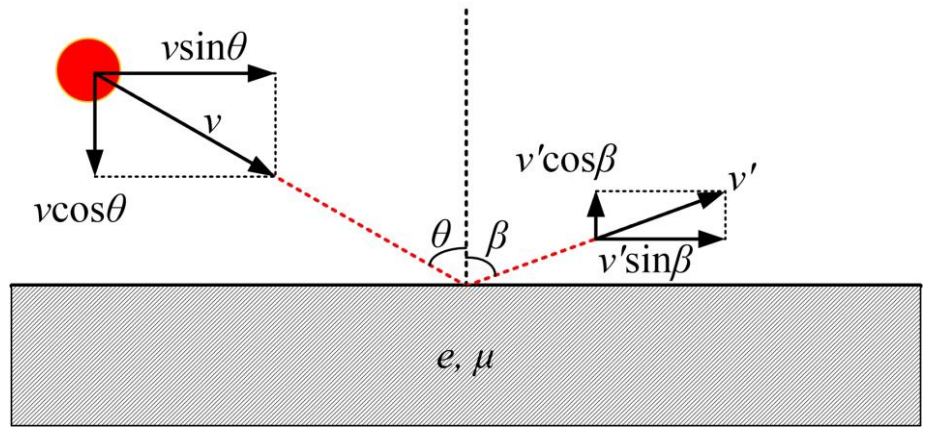

**Fig. 7** - Theoretical collision model used to explain the dependence of maximum FNA $\eta_m$ on the incident angle $\theta$ and kinetic energy $E_k$.

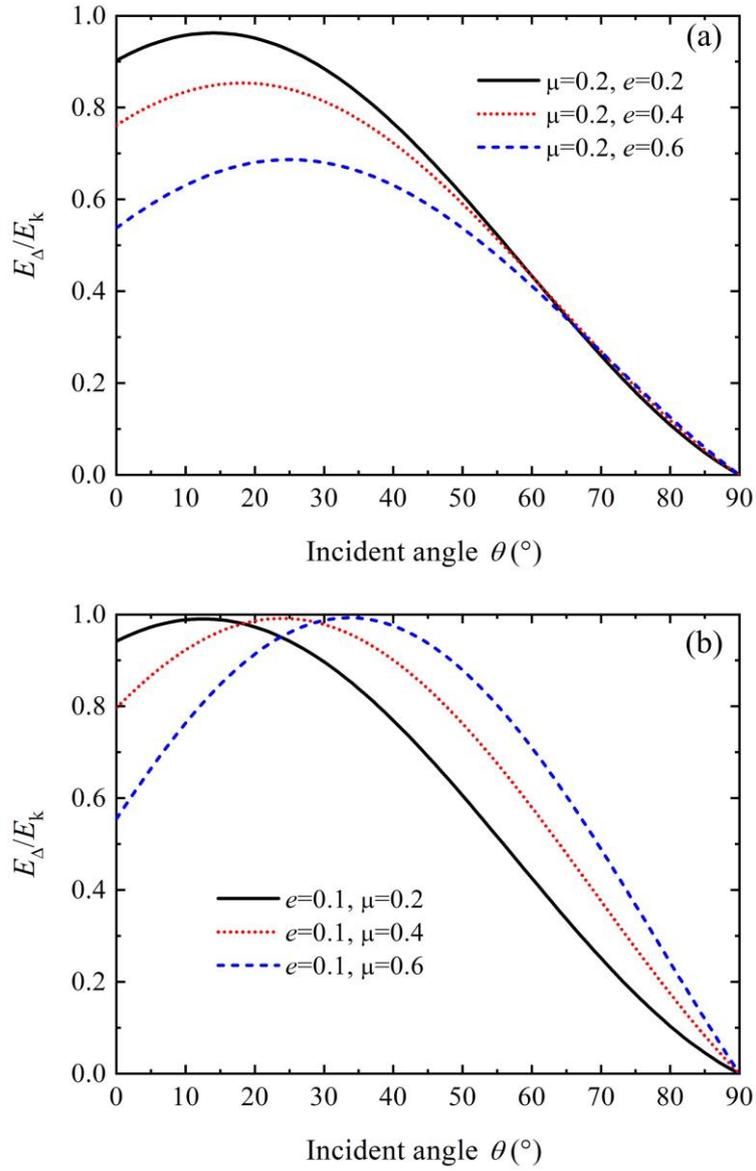

**Fig. 8 -** $E_\Delta / E_k$ as a function of incident angle $\theta$ with various restitution coefficients $e$ and friction coefficients $\mu$.